\documentclass[aps,pre,preprint,showpacs]{revtex4}
\usepackage{bm}
\usepackage{graphicx}
\begin{document}
\title{Hydrodynamic profiles for an impurity in a open vibrated granular gas}
\author{J. Javier Brey}
\email{brey@us.es}
\author{M.J. Ruiz-Montero}
\author{F. Moreno}
\affiliation{F\'{\i}sica Te\'{o}rica, Universidad de Sevilla,
Apdo.\ de Correos 1065, E-41080 Sevilla, Spain}

\date{today}

\begin{abstract}
The hydrodynamic state of an impurity immersed in a low density
granular gas is analyzed. Explicit expressions for the temperature
and density fields of the impurity in terms of the hydrodynamic
fields of the gas are derived. It is shown that the ratio between
the temperatures of the two components, measuring the departure
from energy equipartition, only depends on the mechanical
properties of the particles, being therefore constant in the bulk
of the system. This ratio plays an important role in determining
the density profile of the intruder and its position with respect
to the gas, since it determines the sign of the pressure diffusion
coefficient. The theoretical predictions are compared with
molecular dynamics simulation results for the particular case of
the steady state of an open vibrated granular system in absence of
macroscopic fluxes, and a satisfactory agreement is found.
\end{abstract}


\maketitle

\section{Introduction}
\label{s1} As a consequence of the inelasticity of collisions,
granular fluids exhibit a series of behaviors that are in sharp
contrast with those of molecular fluids. One of them is the
absence of energy equipartition in a granular mixture of
mechanically different species. The granular temperatures of the
components of the mixture, defined from the average kinetic
energies, are different. Although this possibility was already
pointed out many years ago \cite{JyM87}, it has not been until
recently that a systematic study of the effect has been started.
For the homogeneous state of a freely cooling binary mixture of
inelastic spheres, an explicit expression for the ratio of
temperatures of the two components has been obtained from an
approximate solution of the kinetic Enskog equations \cite{GyD99}.
The accuracy of this prediction for weak dissipation and low
density has been confirmed by Molecular Dynamics (MD) simulations
\cite{DHGyD02}.

   The above homogeneous cooling state (HCS) is not accesible
experimentally. In order to maintain a granular system fluidized,
an external energy supply is required. This is often carried out
by means of vibrating walls or external fields, that generate
macroscopic gradients in the system because of inelasticity. At a
theoretical level, {\em homogeneously driven} granular fluids have
also been considered. In these models, stochastic forces injecting
energy are added, then allowing a system otherwise isolated to
reach a steady state. The lack of energy equipartition in the
steady state of homogeneously driven granular mixtures has been
also analyzed \cite{ByT02}, by extending the methods of
\cite{GyD99}. Nevertheless, the relationship between this kind of
driving and actual experiments is uncertain. The non-equipartition
has also been confirmed by MD simulations of simple shear flows
\cite{GDyH05} and of vibrated granular gases (in absence of
gravity) \cite{ByT02a}.

   Experimental evidence of the coexistence of different
temperatures in strongly vibrated granular mixtures has been
reported in both two-dimensional \cite{FyM02} and
three-dimensional \cite{WyP02} systems. Moreover, the dependence
of the temperature ratio on the several parameters characterizing
the system was investigated. Quite surprisingly, it has been
observed that the above ratio remains practically constat in the
bulk of the system, in spite of the rather large gradients
exhibited by each of the partial temperatures.

   The lack of energy equipartition has consequences on the
expressions of the transport coefficients of a granular mixture
\cite{GyD02}. Therefore, it must also affect the shape of the
hydrodynamic profiles and, in particular, the density distribution
of each of the components in inhomogeneous mixtures. Consequently,
it seems clear that the temperature difference can play a role in
the segregation phenomena occurring in vibrated granular mixtures
\cite{DRyC93,BEKyR03}.

   Here, the particular case of a low density
binary mixture in the tracer limit, i.e. when the mole fraction of
one of the components is very low, will be considered. The
simplicity of the system allows a detailed and controlled
discussion of the hydrodynamic profiles of the tracer component,
that can be expressed analytically in terms of those of the other
(excess) component. The starting point are the kinetic Boltzmann
equations for the mixture, and the analysis is based on the
Chapman-Enskog method. The theory is first formulated for a
general kind of states and later on particularized for the steady
state of an open vibrated system. Then, the theoretical
predictions are compared with MD simulation results, and a quite
satisfactory agreement is found over a wide range of values of the
parameters defining the mechanical properties of the system. A
short summary of some of the results presented here was given in
ref. \cite{BRyM05}.

The plan of the paper is as follows. In Sec.\ \ref{s2}, some on
the fundamental ideas on which the Chapmann-Enskog method, as
applied to a one-component system, is based are shortly reviewed.
They are used in Sec.\, \ref{s3}, where an explicit expression for
the ratio between the local temperatures of the two components is
derived. This ratio only depends on the mechanical properties of
the particles, being independent of the gradients of the
hydrodynamic fields. An extensive comparison with MD results for a
vibrated granular system in presence of gravity is carried out.
The shape of the density profile for the tracer component is
investigated in Sec. \ref{s4}, where an expression for the mass
flux relative to the local flow is derived to first order in the
gradients of the mixture fields (Navier-Stokes order). For the
particular case of the steady state of an open vibrated system, an
expression for the density profile of the tracer component in
terms of the fields of the excess component is obtained. Since
analytical forms for these fields are known, the expression can be
integrated numerically. Some details of this are given in the
Appendix. Again, the theoretical predictions compare well with MD
simulation results, although some significant discrepancies show
up as the difference in mass of the particles is increased.
Special emphasis is put on the relationship between the breakdown
of energy equipartition and the relative position of the centers
of mass of the two species. Finally, Sec.\ \ref{s5} contains some
further discussion of the results, as well as some comments on
their applicability to arbitrary states.

\section{Chapman-Enskog description of a granular gas}
\label{s2} As already said, the aim of this work is to analyze
some properties of the hydrodynamic profiles of a dilute granular
binary mixture in the tracer limit, i.e. when the mole fraction of
one of the species is very small. More precisely, the emphasis is
put on the relationship between the profiles of the tracer
component  and those of the mixture in an arbitrary state. It is
assumed that the concentration of the tracer component is so small
that its presence does not affect the state of the granular
mixture that is, therefore, determined by the state of the other
(excess) component. At a kinetic theory level, this implies that
the evolution of the one-particle distribution function of the
excess component obeys a closed nonlinear Boltzmann equation.
Moreover, in the evolution of the distribution function of the
tracer component, the mutual interactions between their particles
can be neglected, as compared with the interactions with the
particles of the excess component. Consequently, its distribution
obeys a linear Boltzmann-Lorentz equation. This is formally
equivalent to consider an impurity or intruder immersed in a
dilute granular gas, and this will be the terminology used in the
following. Let us start by shortly reviewing some basic aspects of
the Chapman-Enskog method applied to a one-component granular gas
that will be needed in the following.

Consider a low density gas composed by smooth inelastic hard
spheres ($d=3$) or disks ($d=2$) of mass $m$ and diameter
$\sigma$. The inelasticity of collisions is modelled by a
velocity-independent coefficient of normal restitution $\alpha$,
defined in the interval $0 < \alpha \leq 1$. The particles are
submitted to an external force of the gravitational type, so the
force acting on each particle has the form ${\bm
F}=-mg_{0}\widehat{\bm e}_{z}$, where $g_{0}$ is a constant and
$\widehat{\bm e}_{z}$ the unit vector in the positive direction of
the $z$ axis.
      It is assumed that in the low density limit, the time
evolution of the one-particle distribution function of the gas,
$f({\bm r},{\bm v},t)$, is accurately described by the nonlinear
inelastic Boltzmann equation
\begin{equation}
\label{2.1} \left( \partial_{t}+{\bm v} \cdot \nabla-g_{0}
\frac{\partial}{\partial v_{z}} \right) f({\bm r},{\bm v},t)=
J[{\bm v}|f,f],
\end{equation}
where $J$ is the Boltzmann collision operator describing the
scattering of pairs of particles,
\begin{eqnarray}
\label{2.2} J[{\bm v}|f,f] & = &  \sigma^{d-1} \int d{\bm v}_{1}
\int d\widehat{\bm \sigma}\, \Theta ({\bm g} \cdot \widehat{\bm
\sigma}) \nonumber \\
&& \times {\bm g} \cdot \widehat{\bm \sigma} \left[\alpha^{-2}
f({\bm r},{\bm v}^{\prime},t)f({\bm r},{\bm v}^{\prime}_{1},t)
\right. \nonumber \\
&& \left. -f({\bm r},{\bm v},t)f({\bm r},{\bm v}_{1},t) \right].
\end{eqnarray}
Here $\Theta$ is the Heaviside step function, $\widehat{\bm
\sigma}$ is the unit vector pointing from the center of particle
$1$ to the center of the other colliding particle at contact, and
${\bm g}={\bm v}-{\bm v}_{1}$ the relative velocity. Moreover,
${\bm v}^{\prime}$ and ${\bm v}^{\prime}_{1}$ are the
precollisional or restituting velocities, i.e. the initial values
of the velocities leading to ${\bm v}$ and ${\bm v}_{1}$ following
the binary collision defined by $\widehat{\bm \sigma}$,
\begin{eqnarray}
\label{2.3} {\bm v}^{\prime} &=&{\bm v}-\frac{1+\alpha}{2 \alpha}
({\bm g} \cdot \widehat{\bm \sigma}) \widehat{\bm \sigma},
\nonumber \\
{\bm v}^{\prime}_{1} &=&{\bm v}_{1}+\frac{1+\alpha}{2 \alpha}
({\bm g} \cdot \widehat{\bm \sigma}) \widehat{\bm \sigma}.
\end{eqnarray}
Macroscopic fields number density, $n({\bm r},t)$, flow velocity
${\bm u}({\bm r},t)$, and granular temperature, $T({\bm r},t)$,
are defined in the usual way as velocity moments of the
distribution function,
\[
n({\bm r},t)= \int d{\bm v}\, f({\bm r},{\bm v},t),
\]
\[
n({\bm r},t) {\bm u}({\bm r},t)= \int d{\bm v}\, {\bm v} f({\bm
r},{\bm v},t),
\]
\begin{equation}
\label{2.4} n({\bm r},t)T({\bm r},t)= \int d{\bm v}
\frac{mV^{2}}{d} f({\bm r},{\bm v},t),
\end{equation}
where ${\bm V}={\bm v}-{\bm u}$ is the peculiar velocity. Balance
equations for the above fields are directly obtained from the
Boltzmann equation,
\begin{equation}
\label{2.5}
\partial_{t}n+\nabla \cdot (n {\bm u})=0,
\end{equation}
\begin{equation}
\label{2.6}
\partial_{t} {\bm u} +{\bm u} \cdot \nabla {\bm u} +(mn)^{-1}
\nabla \cdot {\sf P} +g_{0} \widehat{\bm e}_{z}=0,
\end{equation}
\begin{equation}
\label{2.7}
\partial_{t} T+{\bm u} \cdot \nabla T+ 2 (nd)^{-1} (\nabla \cdot
{\bm q}+{\sf P}: \nabla {\bm u})+T \zeta=0.
\end{equation}
The pressure tensor ${\sf P}$ and the heat flux ${\bm q}$ are
given by
\begin{equation}
\label{2.8} {\sf P}= \int d{\bm v}\, m {\bm V}{\bm V} f({\bm
r},{\bm v},t),
\end{equation}
\begin{equation}
\label{2.9} {\bm q}=\int d{\bm v}\, \frac{m}{2} V^{2}{\bm V}
f({\bm r},{\bm v},t),
\end{equation}
respectively. The cooling rate, $\zeta$, in Eq. (\ref{2.7}) is due
to the energy dissipation in collisions and it is also a
functional of the distribution function,
\begin{equation}
\label{2.10} \zeta({\bm r},t)=-(nTd)^{-1} \int d{\bm v}\, m v^{2}
J[{\bm v}|f,f].
\end{equation}
The balance equations (\ref{2.5})-(\ref{2.7}) only become a set
closed hydrodynamic equations for the fields once ${\sf P}$, ${\bm
q}$ and $\zeta$ are expressed as functionals of them. This
requires to find a solution to the Boltzmann equation such that
all the space and time dependence occurs through $n$, ${\bm u}$,
and $T$. In the Chapman-Enskog procedure, this {\em normal}
solution is generated by expressing it as a series expansion in a
formal non-uniformity parameter $\epsilon$,
\begin{equation} \label{2.11} f= f^{(0)}+\epsilon
f^{(1)}+\epsilon^{2} f^{(2)}+ \ldots,
\end{equation}
where each factor of $\epsilon$ means an implicit gradient of a
macroscopic field. Use of the above expansion in the definitions
of the fluxes and the cooling rate gives a corresponding expansion
for them. Finally, the time derivatives of the fields are also
expanded in the gradients,
$\partial_{t}=\partial_{t}^{(0)}+\epsilon
\partial_{t}^{(1)}+\ldots$, by means of the balance equations. Then, it is
\begin{equation}
\label{2.12} \partial_{t}^{(0)}n=0, \quad \partial_{t}^{(0)}{\bm
u}=0, \quad \partial_{t}^{(0)} T=-T \zeta^{(0)},
\end{equation}
where the zeroth order cooling rate $\zeta^{(0)}$ is given by
\begin{equation}
\label{2.13} \zeta^{(0)}({\bm r},t)=-(nTd)^{-1} \int d{\bm v}\,
mV^{2} J[{\bm v}|f^{(0)},f^{(0)}].
\end{equation}
In this way, to zeroth order in $\epsilon$ it is obtained:
\begin{equation}
\label{2.14} -\zeta^{(0)} T \frac{\partial}{\partial T}
f^{(0)}=J[{\bm v}|f^{(0)},f^{(0)}].
\end{equation}
The solution to this equation is chosen such that it verifies the
conditions
\[
\int d{\bm v}\, f^{(0)}({\bm r},{\bm v},t)=n({\bm r},t),
\]
\[
\int d{\bm v}\, {\bm v} f^{(0)}({\bm r},{\bm v},t)=n({\bm r},t)
{\bm u}({\bm r},t),
\]
\begin{equation}
\label{2.15} \int d{\bm v} \frac{mV^{2}}{d} f^{(0)}({\bm r},{\bm
v},t)= n({\bm r},t) T({\bm r},t),
\end{equation}
i.e., it leads to the same hydrodynamic fields as the complete
distribution function. This choice is consistent with the
solubility conditions for the set of equations generated by the
Chapman-Enskog method \cite{FyK72}. Thus $f^{(0)}$ is easily
generated from the distribution function for the homogeneous
cooling state (HCS) of a dilute gas \cite{GyS95} by substituting
the homogeneous density and temperature fields by the local values
$n({\bm r},t)$ and $T({\bm r},t)$, and replacing the velocity
${\bm v}$ by the peculiar one, ${\bm V}({\bm r},t)$. Therefore, it
has the form \cite{GyS95}
\begin{equation}
\label{2.16} f^{(0)}({\bm r},{\bm v},t)=n v_{T}^{-d}(t) \chi
(V/v_{T}),
\end{equation}
where $v_{T}^{2}({\bf r},t)=2 T({\bm r},t)/m$ is a local thermal
velocity. As a consequence, Eq. (\ref{2.14}) can be rewritten as
\begin{equation}
\label{2.17} \frac{\zeta^{(0)}}{2} \frac{\partial}{\partial {\bm
V}} \cdot \left( {\bm V} f^{(0)} \right) =J[{\bm
v}|f^{(0)},f^{(0)}].
\end{equation}
Using Eq.\ (\ref{2.16}) it is easily obtained:
\begin{equation}
\label{2.18}
\partial_{t}^{(1)}n=-\nabla \cdot (n{\bm u}),
\end{equation}
\begin{equation}
\label{2.19}
\partial_{t}^{(1)}{\bm u}=-{\bm u}\cdot \nabla{\bm u}-(mn)^{-1}
\nabla p-g_{0} \widehat{\bm e}_{z},
\end{equation}
\begin{equation}
\label{2.20}
\partial_{t}^{(1)}T=-{\bm u} \cdot \nabla T-\frac{2T}{d} \nabla \cdot
{\bm u}.
\end{equation}
In the last expression, use has been made of the fact that
$\zeta^{(1)}=0$ because of symmetry considerations \cite{BDKyS98}.
Moreover, $p=nT$ is the pressure of the gas. The procedure can be
continued and Navier-Stokes hydrodynamic equations for a dilute
granular gas, with explicit expressions for the transport
coefficients in the first Sonine approximation, have been derived
\cite{BDKyS98,ByC01}.

\section{Temperature of the impurity}
\label{s3} Let us suppose now that an impurity or intruder of mass
$m_{0}$ and diameter $\sigma_{0}$ is added to the system. It is
assumed that the presence of the intruder does not affect the
state of the gas, so that the one-particle distribution function
of the gas particles is still determined by the nonlinear
Boltzmann equation (\ref{2.1}). Moreover, the macroscopic flow
velocity and temperature for the mixture formed by the gas plus
the intruder are the same as those for the gas alone, i.e. they
are given by Eqs.\ (\ref{2.4}).

The distribution function of the intruder, $f_{0}({\bm r},{\bm
v},t)$, obeys the linear Boltzmann-Lorentz equation
\begin{equation}
\label{2.21} \left( \partial_{t}+{\bm v} \cdot \nabla-g_{0}
\frac{\partial}{\partial v_{z}} \right) f_{0}({\bm r},{\bm v},t)=
J_{0}[{\bm v}|f_{0},f],
\end{equation}
where the collision operator now is
\begin{eqnarray}
\label{2.22} J_{0}[{\bm v}|f_{0},f]& = & \overline{\sigma}^{d-1}
\int d{\bm v}_{1}\, \int d\widehat{\bm \sigma}\, \Theta ({\bm g}
\cdot \widehat{\bm \sigma}) \nonumber \\
&  & \times  {\bm g} \cdot \widehat{\bm \sigma}
\left[\alpha_{0}^{-2} f_{0}({\bm r},{\bm v}^{\prime
\prime},t)f({\bm r},{\bm v}^{\prime \prime }_{1},t) \right.
\nonumber \\
&& \left. -f_{0}({\bm r},{\bm v},t)f({\bm r},{\bm v}_{1},t)
\right].
\end{eqnarray}
Here, $\overline{\sigma}=(\sigma + \sigma_{0})/2$ and $\alpha_{0}$
is the coefficient of normal restitution for impurity-gas
collisions. The precollisional velocities in this case are given
by
\begin{eqnarray}
\label{2.23} {\bm v}^{\prime \prime} & = & {\bm
v}-\frac{(1+\alpha_{0})\Delta}{\alpha_{0}} ({\bm g} \cdot
\widehat{\bm \sigma}) \widehat{\bm \sigma}, \nonumber \\
{\bm v}^{\prime \prime}_{1} & = & {\bm
v}_{1}+\frac{(1+\alpha_{0})(1-\Delta)}{\alpha_{0}} ({\bm g} \cdot
\widehat{\bm \sigma}) \widehat{\bm \sigma},
\end{eqnarray}
with $\Delta=m/(m+m_{0})$. The number density for the intruder is
\begin{equation}
\label{2.24} n_{0}({\bm r},t)= \int d{\bm v}\, f_{0}({\bm r},{\bm
v},t).
\end{equation}

The Chapman-Enskog procedure when applied to a mixture, assumes
the existence of a normal solution of the Boltzmann equations for
the mixture in which all the space and time dependence of the one
particle distribution function of each of the species occurs
through a functional dependence on the hydrodynamic fields of the
mixture \cite{FyK72}. These fields can be chosen in different,
equivalent ways. Here they are taken to be the concentration of
the impurity $x_{0}=n_{0}/(n_{0}+n) \rightarrow n_{0}/n$, the
pressure $p$, the local flow velocity ${\bm u}$, and the
temperature $T$, i.e.
\begin{equation}
\label{2.27} f_{0}({\bm r},{\bm v},t)=f_{0}[{\bm v}|x_{0}({\bm
r},t),p({\bm r},t),{\bm u}({\bm r},t),T({\bm r},t)].
\end{equation}
Then, an expansion similar to the one given in Eq.\ (\ref{2.11})
is considered for the distribution function of the impurity,
$f_{0}=f_{0}^{(0)}+\epsilon f_{0}^{(1)}+\epsilon^{2}
f_{0}^{(2)}+\ldots$. To lowest order in $\epsilon$, Eq.
(\ref{2.21}) becomes
\begin{equation}
\label{2.28} - \zeta^{(0)} \left( p\frac{\partial}{\partial p}+T
\frac{\partial}{\partial T} \right) f_{0}^{(0)}=J_{0}[{\bm
v}|f_{0}^{(0)},f^{(0)}].
\end{equation}
Taking into account that dimensional analysis requires that
$f_{0}^{(0)}$ is of the form
\begin{equation}
\label{2.29} f_{0}^{(0)}=x_{0} \frac{p}{T}\, v_{T}^{-d} \chi_{0}
\left( V/v_{T}\right),
\end{equation}
Eq.\ (\ref{2.28}) is seen to be equivalent to
\begin{equation}
\label{2.30} \frac{\zeta^{(0)}}{2} \frac{\partial}{\partial {\bm
V}} \cdot \left( {\bm V} f_{0}^{(0)} \right)=J_{0}[{\bm
v}|f_{0}^{(0)},f^{(0)}].
\end{equation}
The solution of this equation is chosen such that
\begin{equation}
\label{2.31} \int d{\bm v}\, f_{0}^{(0)}({\bm r},{\bm v},t)=\int
d{\bm v} f_{0}({\bm r},{\bm v},t)=n_{0}({\bm r},t),
\end{equation}
\begin{equation}
\label{2.32} \int d{\bm v}\ {\bm v} f_{0}^{(0)} ({\bm r},{\bm
v},t) = n_{0}({\bm r},t) {\bm u}({\bm r},t),
\end{equation}
\begin{eqnarray}
\label{2.33} \int d{\bm v}\,  m V^{2} f^{(0)}_{0}({\bm r},{\bm
v},t)& = & \int d{\bm v}\, m V^{2} f_{0}({\bm r},{\bm v},t)
\nonumber \\
&=& d n_{0}({\bm r},t) T_{0}({\bm r},t).
\end{eqnarray}
The last equality in Eq. (\ref{2.33}) defines the local
temperature, $T_{0}$, of the intruder. The above requirements are
consistent with the solubility conditions of the equations
generated by the Chapman-Enskog method. It is worth to stress that
Eq.\ (\ref{2.30}) holds with independence of the specific form of
the hydrodynamic fields of the mixture. Thus multiplication of
that equation by $m_{0}V^{2}$ and integration over ${\bm V}$
yields
\begin{equation}
\label{2.34} \zeta^{(0)}({\bm r},t) =\zeta_{0}^{(0)}({\bm r},t),
\end{equation}
where
\begin{equation}
\label{2.35} \zeta_{0}^{(0)}({\bm r},t)=-(n_{0}T_{0}d)^{-1} \int
d{\bm V}\, m_{0} V^{2} J_{0}[{\bm v}|f_{0}^{(0)},f^{(0)}]
\end{equation}
is the lowest order in the gradients of the cooling rate for the
temperature of the intruder.

The evaluation of the cooling rates $\zeta^{(0)}$ and
$\zeta_{0}^{(0)}$ requires to solve Eqs.\ (\ref{2.14}) and
(\ref{2.30}). This can be done in a systematic way by expanding
$f^{(0)}$ and $f_{0}^{(0)}$ in terms of an ensemble of orthogonal
polynomials \cite{GyD99}. Here we will consider a leading order
approximation that is expected to give quite accurate results, at
least for not very strong inelasticity. The zeroth order
distributions are approximated by Gaussians,
\begin{equation}
\label{2.36} f^{(0)}({\bm r},{\bm v},t)= n \left(\frac{2\pi T}{m}
\right)^{-d/2} e^{-mV^{2}/2T},
\end{equation}
\begin{equation}
\label{2.37} f^{(0)}_{0}({\bm r},{\bm v},t)= n_{0}\left( \frac{2
\pi T_{0}}{m_{0}} \right)^{- d/2} e^{-m_{0}V^{2}/2T_{0}}.
\end{equation}
Note that these expressions are consistent with the conditions
(\ref{2.14}) and (\ref{2.31})-(\ref{2.33}). Employing them, it is
straightforward to calculate the cooling rates. The technical
details needed to evaluate the integrals have already been
discussed several times \cite{GyD99,ByT02,DByL02}, and they will
not be reproduced here. The result is
\begin{equation}
\label{2.38} \zeta^{(0)*} \equiv \frac{\zeta^{(0)}({\bm
r},t)}{n({\bm r},t) \sigma^{d-1} v_{T}(t)} =\frac{\sqrt{2} \pi
^{(d-1)/2}}{\Gamma (d/2)d}\, (1-\alpha^{2}),
\end{equation}
\begin{equation}
\label{2.39} \zeta_{0}^{(0)*} \equiv \frac{\zeta_{0}^{(0)}({\bm
r},t)}{n({\bm r},t) \sigma^{d-1} v_{T}(t)} = \nu^{*}_{0}
(1+\phi)^{1/2} \left(1-h\frac{1+\phi}{\phi} \right),
\end{equation}
where $h=m(1+\alpha_{0})/2(m+m_{0})$, $\nu^{*}_{0}$ is a
dimensionless collision rate,
\begin{equation}
\label{2.40} \nu^{*}_{0}=\frac{8h \pi^{(d-1)/2}}{\Gamma(d/2)d}
\left( \frac{\overline{\sigma}}{\sigma} \right)^{d-1},
\end{equation}
and $\phi$ is the ratio of the mean square velocities for the
intruder and fluid particles,
\begin{equation}
\label{2.41} \phi=\frac{ m T_{0}(z)}{m_{0} T(z)}.
\end{equation}
Substitution of Eqs. (\ref{2.38}) and (\ref{2.39}) into Eq.\
(\ref{2.34}) gives
\begin{equation}
\label{2.42}  (1+\phi)^{1/2} \left( 1-h \frac{1+\phi}{\phi}
\right) = \frac{\beta}{h},
\end{equation}
where the parameter $\beta$ is given by
\begin{equation}
\label{2.43} \beta \equiv  \frac{1-\alpha^{2}}{4 \sqrt{2}} \left(
\frac{\sigma}{\overline{\sigma}} \right)^{d-1}.
\end{equation}
The solution of this equation provides the expression for $
T_{0}({\bm r},t)/T({\bm r},t)$. There is a unique real solution
for all allowed values of $h$ and $\beta$. For elastic collisions
($\alpha=\alpha_{0}=1$), the solution is $\phi=m/m_{0}$ and energy
equipartition follows. If only the intruder-gas collisions are
inelastic ($\alpha=1$), the result $\phi=h/(1-h)$ derived by
Martin and Piasecki for an inelastic impurity in an equilibrium
elastic gas is recovered \cite{MyP99}.

The simplicity of Eq.\ (\ref{2.42}) may appear as a surprise,
specially taking into account its generality. No particular state
of either the gas or the intruder has been considered and,
nevertheless, the temperature ratio is given as a function of only
the mechanical properties of the particles. The existence of an
expression for the temperature ratio is a consequence of the
assumption that there is a hydrodynamic description of the mixture
in terms of only $x_{0}$, $p$, ${\bm u}$, and $T$. Therefore,
although the partial temperatures of the intruder and the gas are
different, they are not needed to specify the macroscopic state of
the mixture. Nevertheless, the theory provides a relationship to
determine them. The fact that Eq.\ (\ref{2.42}) does not involve
any of the other hydrodynamic fields follows from the tracer and
dilute limits we are considering.

The above relation between the partial temperatures is not
modified when applied to the HCS, consistently with previous
results \cite{DByL02,SyD01}. This does not imply by itself any
kind of {\em local HCS approximation}, in the same way as energy
equipartition in molecular systems does not imply local
equilibrium, and it is valid independently of the gradients
present in the system. On the other hand, the above results do not
hold if the characterization of the macroscopic state of the
mixture also requires to specify the partial temperatures of the
gas and the intruder and, therefore, the set of independent
hydrodynamic variables (and also the number of equations) must be
expanded. This seems to be the kind of approximation followed in
\cite{JyM87}.

In order to check the accuracy of the theoretical prediction given
by Eq.\ (\ref{2.42}), we have performed MD simulations with an
event driven algorithm. The particular situation we have
considered is a two-dimensional system confined in a rectangular
box of width $W$, open on the top, and submitted to an external
field of the gravitational type, as defined above Eq.\
(\ref{2.1}). Periodic boundary conditions are enforced in the
direction perpendicular to the field. To maintain the system
fluidized, energy is continuously supplied through the wall
located at $z=0$, which is vibrating. For simplicity, and given
that we are interested in the bulk properties of the system, the
wall moves in a sawtooth way, with very small amplitude and high
frequency, so that it can be considered that all the particles
colliding with the wall find it at $z=0$ and with the same
velocity ${\bm v}_{W}=v_{W} \widehat{\bm e}_{z}$
\cite{McyL98,BRyM01}. The particle-wall collisions are elastic.

   Under the above conditions, it has been proven that a
one-component granular gas reaches a stationary state with only
gradients in the $z$-direction \cite{BRyM01}. Nevertheless, this
state becomes unstable and develops transversal inhomogeneities
when the size of the vibrating wall is larger than a critical
value \cite{SyK01}. In fact, the same happens for a closed system
in absence of gravity \cite{LMyS02,BRMyG02}. In all the
simulations to be reported in the following, the value of $W$ has
been taken small enough so the system stays homogeneous in the
transversal direction.

\begin{figure}
\includegraphics[scale=0.5,angle=-90]{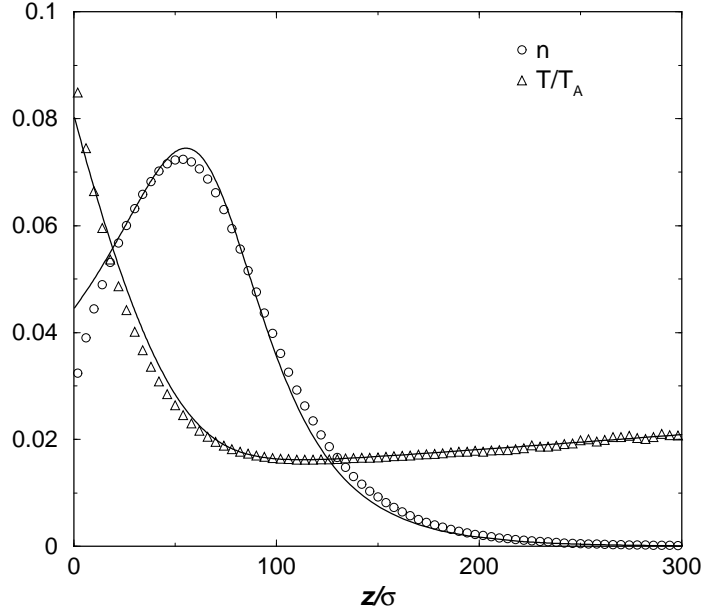}
\caption{Density and temperature vertical profiles for a vibrated
granular gas in the steady state considered in the text. The
number of particles is $N=359$, the restitution coefficient
$\alpha=0.95$, and the size of the vibrating wall $W=50 \sigma$.
The symbols are from  MD simulations and the solid lines the fit
to the hydrodynamic prediction discussed in the Appendix.
\label{fig1}}
\end{figure}

   The above steady state is highly inhomogeneous. An example is
given in Fig \ref{fig1}, where the density and temperature
profiles for a monodisperse system of $N=359$ particles with
$\alpha=0.95$ are shown. In order to plot the temperature profile
on the same scale as the density one, the former has been scaled
with an arbitrary value $T_{A}$. Here and in the following, the
units defined by $m=1$, $\sigma=1$, and $g_{0}=1$ will be used. In
these units, the data in the figure correspond to $W=50$ and
$v_{W}=5$. These values lead to a fluidized state without
transversal inhomogeneities, and with a density low enough as to
expect the Boltzmann equation to give an accurate description of
the system. It is seen that the density profile exhibits a maximum
at which $n \simeq 0.075$. The existence of a density maximum is a
general feature of open vibrated systems as long as the number of
monolayers at rest is large enough \cite{BRyM01}. On the other
hand, the temperature profile presents a minimum which is related
with the existence of a term proportional to the density gradient
in the expression of the heat flux \cite{BRyM01,ByR04}.

\begin{figure}
\includegraphics[scale=0.5,angle=-90]{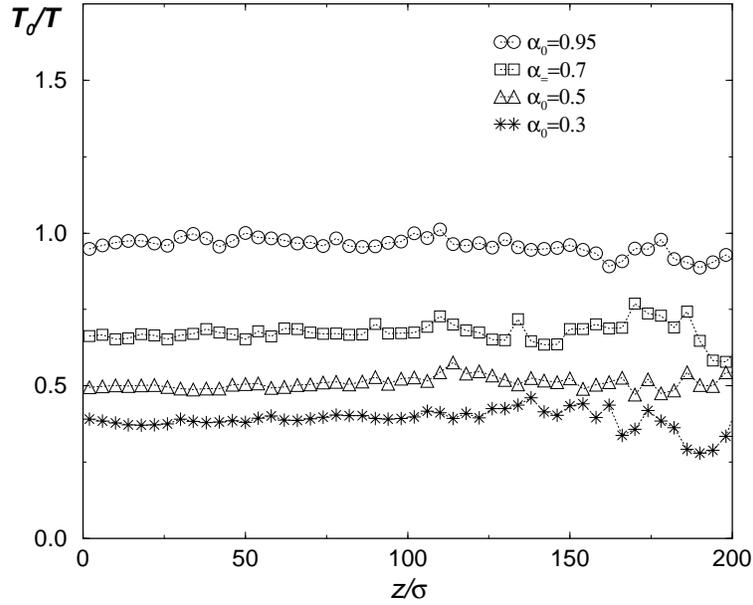}
\caption{Temperature ratio profiles  $T_{0}(z)/T(z)$ for several
values of the restitution coefficient $\alpha_{0}$, as indicated
in the figure. The values of the other parameters are
$\alpha=0.95$, $\sigma_{0}=\sigma$, and $m_{0}=m/2$. \label{fig2}}
\end{figure}

\begin{figure}
\includegraphics[scale=0.5,angle=-90]{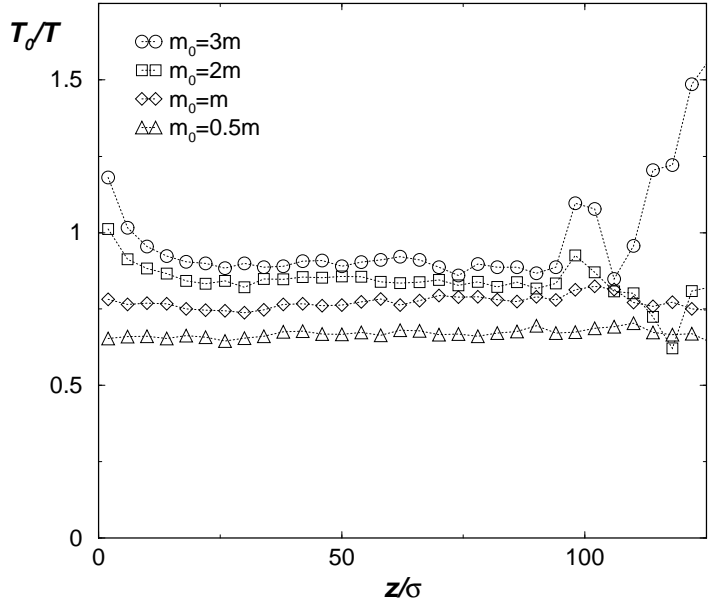}
\caption{Temperature ratio profiles  $ T_{0}(z)/T(z)$ for several
values of the mass ratio $m_{0}/m$, as indicated in the figure.
The values of the other parameters are $\alpha=0.95$,
$\alpha_{0}=0.7$, and  $\sigma_{0}=\sigma$. \label{fig3}}
\end{figure}

\begin{figure}
\includegraphics[scale=0.5,angle=-90]{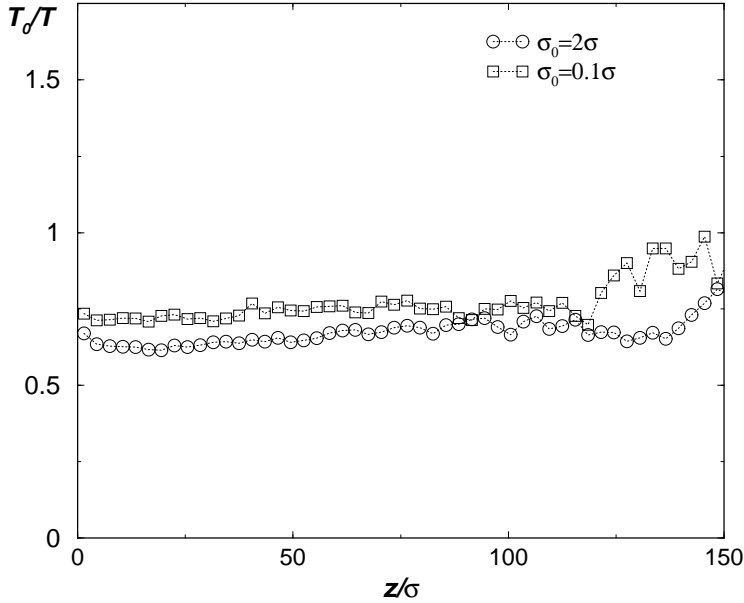}
\caption{Temperature ratio profiles  $ T_{0}(z)/T(z)$ for several
values of the diameter ratio $\sigma_{0}/\sigma$, as indicated in
the figure. The values of the other parameters are $\alpha=0.95$,
$\alpha_{0}=0.6$, and $m_{0}=m$. \label{fig4}}
\end{figure}

Figures \ref{fig2}-\ref{fig4} show the profile of the temperature
ratio $T_{0}(z)/T(z)$ for different values of the mechanical
properties of the particles. It is observed that, aside
statistical fluctuations, the ratio remains practically constant
in a wide region of the system, even if $T$ (and $T_{0}$) varies
significantly. In Fig. \ref{fig2}, the influence of the
inelasticity of the gas-intruder collisions on the temperature
ratio is analyzed. As expected, the deviation from equipartition
increases as $\alpha_{0}$ decreases, becoming more different from
$\alpha$. The influence of the mass ratio $m_{0}/m$ is studied in
Fig. \ref{fig3}, where it is seen that the temperature ratio
significantly deviates from unity when $m_{0}=m$ but $\alpha_{0}$
differs significantly from $\alpha$. This is in contrast with
previous findings in both experiments \cite{FyM02} and MD
simulations \cite{ByT02} of vibrated granular mixtures (not in the
tracer limit), probably because the several coefficients of normal
restitution in those works were too close one another as to
observe this deviation. The figure also shows that the boundary
layer near the vibrating wall, defined as the region in which the
hydrodynamic prediction is not verified, increases as the ratio
$m_{0}/m$ increases. This can be explained as a consequence of the
larger increase of the kinetic energy of the impurity than of the
gas particles when colliding with the vibrating wall. In addition,
the collisions of the intruder with the gas become less efficient
in ``thermalizing'' the former. The same explanation applies to
the strong fluctuations at large heights. Finally, the sensitivity
of $T_{0}/T$ on the diameter ratio is illustrated in Fig.
\ref{fig4}, from where it can be concluded that it is quite weak.

\begin{figure}
\includegraphics[scale=0.5,angle=-90]{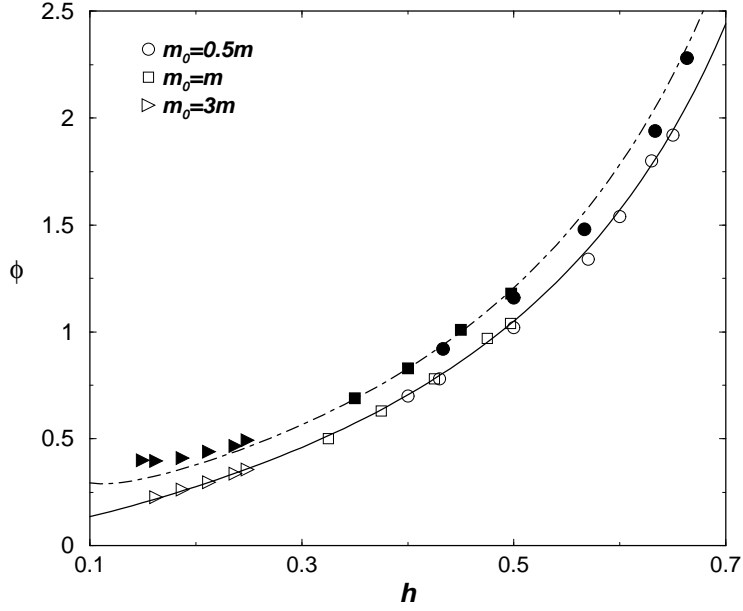}
\caption{Ratio of the mean square velocities $\phi$ as a function
of the dimensionless parameter $h$ defined in the main text. In
all cases $\sigma=\sigma_{0}$. The solid and dashed lines are the
theoretical prediction for $\alpha=0.95$ ($\beta \simeq 0.0172$)
and $\alpha=0.8$ ($\beta \simeq 0.0636$), respectively, while the
empty and filled symbols are from the MD simulations for the same
two values of $\alpha$. The same symbol is used for simulation
results differing only in the value of the restitution coefficient
$\alpha_{0}$. \label{fig5}}
\end{figure}

\begin{figure}
\includegraphics[scale=0.5,angle=-90]{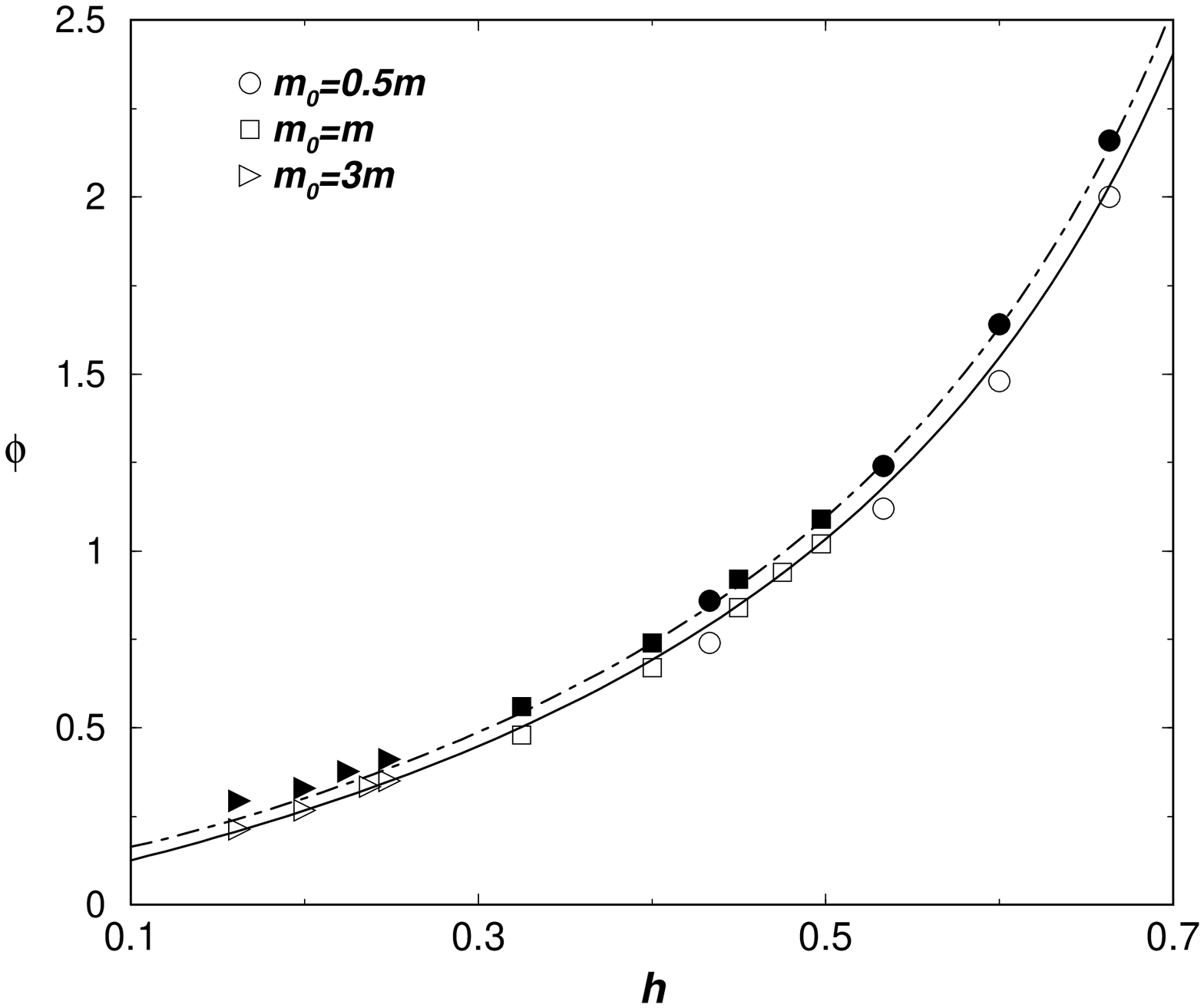}
\caption{Ratio of the mean square velocities $\phi$ as a function
of the dimensionless parameter $h$ defined in the main text. In
all cases $\alpha=0.95$. The solid and dashed lines are the
theoretical prediction for $\sigma_{0}/\sigma=2$ ($\beta \simeq
0.0115$) and $\sigma_{0}/\sigma =0.1$ ($\beta \simeq 0.0313)$,
respectively, while the empty and filled symbols are from the MD
simulations for those two values of the diameter ratio. The same
symbol is used for simulation results differing only in the value
of the restitution coefficient $\alpha_{0}$. \label{fig6}}
\end{figure}

A quantitative comparison between the theoretical prediction given
by Eq. (\ref{2.42}) and the MD results is presented in Figs.
\ref{fig5} and \ref{fig6}. According with Eq.\ (\ref{2.42}), for a
given value of $\beta$, the value of $\phi$ is a function of only
the parameter $h$. Then, the comparison will be presented by
considering series of simulations in which the value of the
coefficient of restitution of the gas, $\alpha$, and the diameter
ratio, $\sigma / \sigma_{0}$, and therefore $\beta$ are kept
constant, while the mass ratio $m_{0}/m$ and the restitution
coefficient $\alpha_{0}$ and, therefore $h$ are changed. In Fig.\
\ref{fig5}, two series of data, one corresponding to
$\sigma_{0}=\sigma$, $\alpha=0.8$ ($\beta=0.0636$) and
$\sigma_{0}=\sigma$, $\alpha=0.95$ ($\beta=0.0172$) are presented.
In each case, several values of the mass ratio have been
considered, as indicated in the figure. In addition, the value of
the coefficient of restitution for gas-intruder collisions has
also been varied in the range $0.2 \leq \alpha_{0} \leq 0.99$. To
keep the figure readable, we have used the same symbols for the
data corresponding to the same mass ratio although different
values of $\alpha_{0}$. A quite good agreement between theory and
simulation is observed, specially for the largest value of
$\alpha$. The small but systematic discrepancies for $\alpha=0.8$
and $m_{0}/m=3$ are probably due to the failure of the
hydrodynamic Navier-Stokes equations to accurately describe the
state of the gas, as discussed in \cite{BRyM01}.

The validity of the theory to describe the dependence on the
diameter ratio is checked in Fig.\ \ref{fig6}. Again, the same
symbol is used for the MD results corresponding to situations
differing only in the value of the coefficient of restitution
$\alpha_{0}$. As before, a weak dependence of the temperature
ratio on the diameter of the particles is observed. It can be
concluded that Eq.\ (\ref{2.42}) provides an accurate description
of the temperature ratio over a wide range of values of the
mechanical parameters characterizing both the gas and the
intruder.

\section{Density profile for the impurity}
\label{s4} An equation for the number density of the impurity
$n_{0}$ is directly obtained from Eq.\ (\ref{2.21}),
\begin{equation}
\label{2.25}
\partial_{t}n_{0}+\nabla \cdot ( n_{0} {\bm u})+m_{0}^{-1} \nabla \cdot {\bm j}_{0}=0,
\end{equation}
where ${\bm j}_{0}({\bm r},t)$ is the mass flux for the impurity
relative to the local flow ${\bm u}({\bm r},t)$,
\begin{equation}
\label{2.26} {\bm j}_{0}({\bm r},t)= m_{0} \int d{\bm v}\, {\bm V}
f_{0}({\bm r},{\bm v},t).
\end{equation}

The isotropy of the function $\chi_{0}$ implies that the zeroth
order of the flux mass for the intruder vanishes, i.e. ${\bm
j}_{0}^{(0)}=0$. Then, we proceed to study the next order, that
requires to determine $f_{0}^{(1)}$. To first order in $\epsilon$,
Eq.\ (\ref{2.21}) reads
\begin{eqnarray}
\label{3.1}
\partial_{t}^{(0)}
f_{0}^{(1)}& - &
J_{0}[f_{0}^{(0)},f^{(1)}]-J_{0}[f_{0}^{(1)},f^{(0)}] \nonumber
\\
& = & -\partial_{t}^{(1)}f_{0}^{(0)}-{\bm v}\cdot \nabla
f_{0}^{(0)}+ g_{0} \frac{\partial}{\partial v_{z}}f_{0}^{(0)}.
\end{eqnarray}
The action of the time derivatives $\partial_{t}^{(0)}$ and
$\partial_{t}^{(1)}$ on the hydrodynamic fields is given by Eqs.\
(\ref{2.12}) and (\ref{2.18})-(\ref{2.20}), respectively. Then,
evaluating the right hand side of Eq. (\ref{3.1}) gives
\[
\partial_{t}^{(0)}
f_{0}^{(1)}-J_{0}[f_{0}^{(0)},f^{(1)}]  -
J_{0}[f_{0}^{(1)},f^{(0)}]
\]
\begin{equation} \label{3.2}
= {\bm A}({\bm V})\cdot \nabla x_{0}+ {\bm B}({\bm V}) \cdot
\nabla p+ {\bm C}({\bm V}) \cdot \nabla T   +  {\sf D}({\bm V}):
\nabla {\bm u},
\end{equation}
where the coefficients of the gradients of the fields on the right
hand side are given by
\begin{equation}
\label{3.3} {\bm A}({\bm V})=-\frac{f_{0}^{(0)}}{x_{0}} {\bm V},
\end{equation}
\begin{equation}
\label{3.4} {\bm B}({\bm V})=-\frac{1}{p} \left( f_{0}^{(0)}{\bm
V}+\frac{T}{m} \frac{\partial f_{0}^{(0)}}{\partial {\bm V}}
\right),
\end{equation}
\begin{equation}
\label{3.5} {\bm C}({\bm V})=\frac{1}{T} \left[
f_{0}^{(0)}+\frac{1}{2} \frac{\partial}{\partial {\bm V}} \cdot
\left( {\bm V} f_{0}^{(0)} \right) \right] {\bm V},
\end{equation}
\begin{equation}
\label{3.6} {\sf D}({\bm V})= {\bm V} \frac{\partial
f_{0}^{(0)}}{\partial {\bm V}}-\frac{1}{d} {\bm V} \cdot
\frac{\partial f_{0}^{(0)}}{\partial {\bm V}} {\sf I}.
\end{equation}
Here, ${\sf I}$ is the unit tensor in $d$ dimensions and $x_{0}$
was defined above Eq.\ (\ref{2.27}). Note that the external field
does not occur in Eq. (\ref{3.1}). This is because of the
particular form for it we are considering (gravitational type). To
proceed, we need the expression of $f^{(1)}$. It has been derived
in ref. \cite{BDKyS98} and reads
\begin{equation}
\label{3.7} f^{(1)}=  \mathcal{B}\cdot \nabla p-\mathcal{C} \cdot
\nabla T+\mathcal{D}: \nabla {\bm u},
\end{equation}
where the coefficients $\mathcal{B}$, $\mathcal{C}$, and
$\mathcal{D}$ are functions of the peculiar velocity ${\bm V}$ and
the hydrodynamic fields. Then, the solution of Eq.\ (\ref{3.2}) is
of the form
\begin{equation}
\label{3.8} f_{0}^{(1)}= \mathcal{A}_{0} \cdot \nabla
x_{0}+\mathcal{B}_{0} \cdot \nabla p+\mathcal{C}_{0} \cdot \nabla
T+\mathcal{D}_{0}  : \nabla {\bm u}.
\end{equation}
Substitution of Eqs.\ (\ref{3.7}) and (\ref{3.8}) into Eq.
(\ref{3.2}) and identifying coefficients of independent gradients
yields
\begin{equation}
\label{3.9} -\zeta^{(0)} \left[ \left( p\frac{\partial}{\partial
p}+T \frac{\partial}{\partial p} \right) \mathcal{A}_{0} \right]
-J_{0}[\mathcal{A}_{0},f^{(0)} ]= {\bm A},
\end{equation}
\[
-\zeta_{0} \left[ \left( p \frac{\partial}{\partial p}+T
\frac{\partial}{\partial T} \right)+2 \right]
\mathcal{B}_{0}-J_{0}[f_{0}^{(0)},\mathcal{B}
]-J_{0}[\mathcal{B}_{0},f^{(0)}]
\]
\begin{equation}
\label{3.10}
= {\bm B}+\frac{T \zeta^{(0)}}{p} \mathcal{C}_{0},
\end{equation}
\[
 -\zeta_{0} \left[ \left(  p \frac{\partial}{\partial p}+T
\frac{\partial}{\partial T} \right) +\frac{1}{2} \right]
\mathcal{C}_{0}-J_{0}[f_{0}^{(1)},\mathcal{C}]-J_{0}[\mathcal{C}_{0},f^{(0)}]
\]
\begin{equation}
\label{3.11}={\bm C}-\frac{p \zeta^{(0)}}{2T} \mathcal{B}_{0},
\end{equation}
\[
-\zeta^{(0)} \left[ \left( p \frac{\partial}{\partial p}+T
\frac{\partial}{\partial T} \right) \mathcal{D}_{0}
\right]-J_{0}[f_{0}^{(0)},\mathcal{D}]-J_{0}[\mathcal{D}_{0},f^{(0)}]
\]
\begin{equation}
\label{3.12}={\bm D}({\bm V}).
\end{equation}
Use of Eq.\ (\ref{3.8}) into Eq. (\ref{2.26}) taking into account
symmetry considerations, gives the expression for the mass flux of
the intruder to first order in the gradients,
\begin{equation}
\label{3.13} {\bm j}_{0}=-m_{0} D \nabla x_{0}-\frac{m}{T} D_{p}
\nabla p-\frac{mn}{T} D^{\prime} \nabla T,
\end{equation}
where
\begin{equation}
\label{3.14} D=-\frac{1}{d} \int d{\bm v} {\bm V} \cdot
\mathcal{A}_{0} \end{equation} is the diffusion coefficient,
\begin{equation}
\label{3.15} D_{p}=-\frac{Tm_{0}}{md} \int d{\bm v} {\bm V} \cdot
\mathcal{B}_{0}
\end{equation}
is the pressure diffusion coefficient, and
\begin{equation}
\label{3.16} D^{\prime}=- \frac{Tm_{0}}{mnd} \int d{\bm v}\, {\bm
V}\cdot \mathcal{C}_{0}
\end{equation}
is the thermal diffusion coefficient.

In the following, the quantities $\mathcal{A}_{0}$,
$\mathcal{B}_{0}$, and $\mathcal{C}_{0}$ will be evaluated in the
first Sonine approximation, i.e. we consider
\begin{equation}
\label{3.17} \left(
\begin{array}{c}
\mathcal{A}_{0} \\
\mathcal{B}_{0} \\
\mathcal{C}_{0}
\end{array}
\right) \simeq f^{(0)}_{0} {\bm V} \left(
\begin{array}{c}
a_{0} \\
b_{0} \\
c_{0}
\end{array}
\right),
\end{equation}
with $f^{(0)}_{0}$ being the Maxwellian given by Eq.\
(\ref{2.37}). Consistently, $\mathcal{B}$ and $\mathcal{C}$ are
approximated by
\begin{equation}
\label{3.18} \left(
\begin{array}{c}
\mathcal{B} \\
\mathcal{C}
\end{array}
\right) \simeq f^{(0)} {\bm V} \left(
\begin{array}{c}
b \\
c
\end{array}
\right),
\end{equation}
and it happens that both $b$ and $c$ are zero (see Appendix C in
ref. \cite{BDKyS98}). Substitution of the above expressions into
Eqs. (\ref{3.9})-(\ref{3.11}) gives a set of closed equations for
$a_{0}$, $b_{0}$, and $c_{0}$. Multiplication of these equations
by $m_{0}{\bm V}$ and integration over ${\bm V}$ yields
\begin{equation}
\label{3.19} \left[-\zeta^{(0)} \left(p \frac{\partial}{\partial
p}+T \frac{\partial}{\partial T} \right)+\nu \right]
a_{0}n_{0}T_{0}=-\frac{n_{0}T_{0}}{x_{0}},
\end{equation}
\[
\left[-\zeta^{(0)} \left( p\frac{\partial}{\partial p}+T
\frac{\partial}{\partial T} \right)-2\zeta^{(0)}+\nu \right]
b_{0}n_{0}T_{0}
\]
\begin{equation}
\label{3.20} = -\frac{n_{0}T_{0}(\phi-1)}{p \phi}
+\frac{T\zeta^{(0)} c_{0}n_{0}T_{0}}{p},
\end{equation}
\[
\left[ -\zeta^{(0)} \left(p \frac{\partial}{\partial p}+T
\frac{\partial}{\partial T} \right)-\frac{\zeta^{(0)}}{2} +\nu
\right] c_{0}n_{0}T_{0}
\]
\begin{equation}
 \label{3.21}=-\frac{p \zeta^{(0)}
b_{0}n_{0}T_{0}}{2T},
\end{equation}
where
\begin{equation}
\label{3.22} \nu=-\frac{m_{0}}{n_{0}T_{0}d}\ \int d{\bm V} {\bm V}
\cdot J_{0}[{\bm V}f_{0}^{(0)},f^{(0)}]
\end{equation}
is a collision frequency. It can be easily evaluated by using
standard integration techniques with the result
\begin{equation}
\label{3.23}
\nu=\frac{\nu_{e}(1+\alpha_{0})}{2}(1-\Delta)^{1/2}(1+\phi)^{1/2}
\end{equation}
where
\begin{equation}
\label{3.24} \nu_{e}=\frac{4\sqrt{2} \pi^{(d-1)/2}}{\Gamma(d/2)d}
\overline{\sigma}^{d-1}n \left( \frac{T}{m_{0}} \right)^{1/2}
\Delta^{1/2}
\end{equation}is the elastic limit. An expression for $\nu$
for the case $d=3$, valid for arbitrary concentrations of the two
components of the mixture and incorporating some non-gaussian
contributions, has been obtained in \cite{GyD02}. Eq.\
(\ref{3.23}) is consistent with that result.

>From dimensional analysis, it follows that $a_{0}n_{0}T_{0}
\propto T^{1/2}$, $b_{0}n_{0}T_{0} \propto T^{1/2}/p$, and
$c_{0}n_{0}T_{0} \propto T^{-1/2}$. Then, the temperature and
pressure derivatives in Eqs. (\ref{3.19})-(\ref{3.21}) can be
evaluated and the equations become
\begin{equation}
\label{3.25} \left(\nu-\frac{\zeta^{(0)}}{2} \right)
a_{0}=-x_{0}^{-1},
\end{equation}
\begin{equation}
\label{3.26} \left( \nu -\frac{3 \zeta^{(0)}}{2} \right)
b_{0}=-\frac{(\phi-1)}{p \phi}+\frac{T \zeta^{(0)}}{p}\, c_{0},
\end{equation}
\begin{equation}
\label{3.27} c_{0}=-\frac{p \zeta^{(0)}}{2T \nu}\, b_{0}.
\end{equation}
Therefore,
\begin{equation}
\label{3.28} a_{0}=-\left[ x_{0} \left( \nu-\frac{\zeta^{(0)}}{2}
\right) \right]^{-1},
\end{equation}
\begin{equation}
\label{3.29} b_{0}=- \left[ \frac{p \phi}{\phi-1} \left(
\nu-\frac{3 \zeta^{(0)}}{2}+\frac{\zeta^{(0)^2}}{2\nu} \right)
\right]^{-1},
\end{equation}
\begin{equation}
\label{3.30} c_{0}=-\frac{p \zeta^{(0)}}{2 T \nu}\, b_{0}.
\end{equation}
The expressions of the transport coefficients follow by using the
above results into Eqs.\ (\ref{3.14})-(\ref{3.16}),
\begin{equation}
\label{3.31} D=\frac{nT_{0}}{m_{0}} \left(
\nu-\frac{\zeta^{(0)}}{2} \right)^{-1},
\end{equation}
\begin{equation}
\label{3.32} D_{p}=\frac{n_{0}T_{0}(\phi-1)}{mn \phi} \left( \nu -
\frac{3 \zeta^{(0)}}{2} +\frac{\zeta^{(0)2}}{2 \nu} \right)^{-1},
\end{equation}
\begin{equation}
\label{3.33} D^{\prime}=-\frac{\zeta^{(0)}}{2 \nu}\, D_{p}.
\end{equation}
These expressions are consistent with those reported in ref.
\cite{GyD02} for $d=3$ and without external field.

Let us now particularize the above results for the steady state
considered in the previous Section, i. e. an open vibrated system
with no macroscopic flow and  gradients only in the $z$ direction.
In this case,  Eq.\ (\ref{2.25}) implies that $j_{z}=constant=0$,
since the intruder mass flux must vanish for $z \rightarrow
\infty$. Thus Eqs.\ (\ref{3.13}) and (\ref{3.31})-(\ref{3.33})
yield
\begin{eqnarray}
\label{3.34} \left( 1-\frac{\zeta^{(0)}}{2 \nu} \right)^{-1}
\frac{\partial \ln x_{0}}{\partial z} & =& -\frac{\phi-1}{\phi}
\left( 1-\frac{3\zeta^{(0)}}{2\nu}+ \frac{\zeta^{(0)2}}{2 \nu^{2}}
\right)^{-1}  \nonumber \\
& & \times \left(\frac{\partial \ln p}{\partial z}-
\frac{\zeta^{(0)}}{2 \nu} \frac{\partial \ln T}{\partial z}
\right).
\end{eqnarray}
Using Eqs.\ (\ref{2.38}) and (\ref{3.23}), and also the equation
determining $\phi$, Eq.\ (\ref{2.42}), it is found
\begin{equation}
\label{3.35} \frac{\zeta^{(0)}}{\nu}=2 \left[
1-\frac{h(1+\phi)}{\phi} \right].
\end{equation}
This shows that $\zeta^{(0)}/\nu$ is a function of only $\beta$
and $h$, i.e. of the mechanical properties of the particles
($\phi$ is given in terms of them by Eq.\ (\ref{2.42})), but it
does not depend on the hydrodynamic fields. Therefore, Eq.\
(\ref{3.34}) can be expressed in compact form as
\begin{equation}
\label{3.36} \frac{\partial \ln x_{0}}{\partial z}=- \lambda
(h,\beta) g(z),
\end{equation}
where
\begin{equation}
\label{3.37} \lambda (h, \beta)= \frac{\phi-1}{\phi} \left( 1-
\frac{\zeta^{(0)}}{2 \nu} \right)\left(
1-\frac{3\zeta^{(0)}}{2\nu}+ \frac{\zeta^{(0)2}}{2 \nu^{2}}
\right)^{-1}
\end{equation}
and
\begin{equation}
\label{3.38} g(z)=\frac{\partial \ln p}{\partial
z}-\frac{\zeta^{(0)}}{2 \nu} \frac{\partial \ln T}{\partial z}\, .
\end{equation}
Integration of Eq. (\ref{3.36}) directly provides an expression
for the probability density $P_{0}(z)$ of finding the impurity at
height $z$,
\begin{equation}
\label{3.39} P_{0}(z)= C n(z) \exp \left(- \lambda \int_{0}^{z}
dz^{\prime} g(z^{\prime}) \right),
\end{equation}
with $C$ being the normalization constant.

The hydrodynamic profiles for the steady state of an open vibrated
dilute gas (without impurity) where studied in \cite{BRyM01} and
the results are shorty summarized in the Appendix. There, it is
also indicated how the numerical evaluation of the integral
appearing on the right hand side of Eq.\ (\ref{3.39}) is
implemented.

\begin{figure}
\includegraphics[scale=0.5,angle=-90]{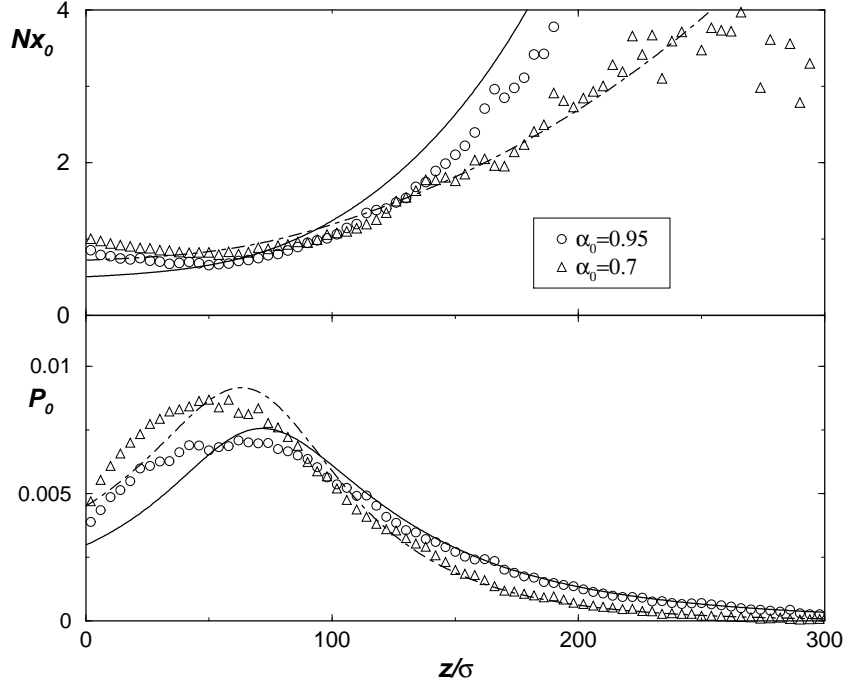}
\caption{Normalized molar fraction $Nx_{0}$ (upper plot) and
density profile $P_{0}(z)$ of the impurity (lower plot) for two
different values of $\alpha_{0}$, as indicated in the figure. The
symbols are from MD simulations, and the solid (dashed) line the
theoretical prediction for $\alpha_{0}=0.95$ ($0.7$) discussed in
the text. The parameter of the gas are: $\alpha=0.95$, $N=359$,
and $W=50 \sigma$. The other parameters of the impurity are
$m_{0}=m/2$ and $\sigma_{0}=\sigma$. \label{fig7}}
\end{figure}

\begin{figure}
\includegraphics[scale=0.5,angle=-90]{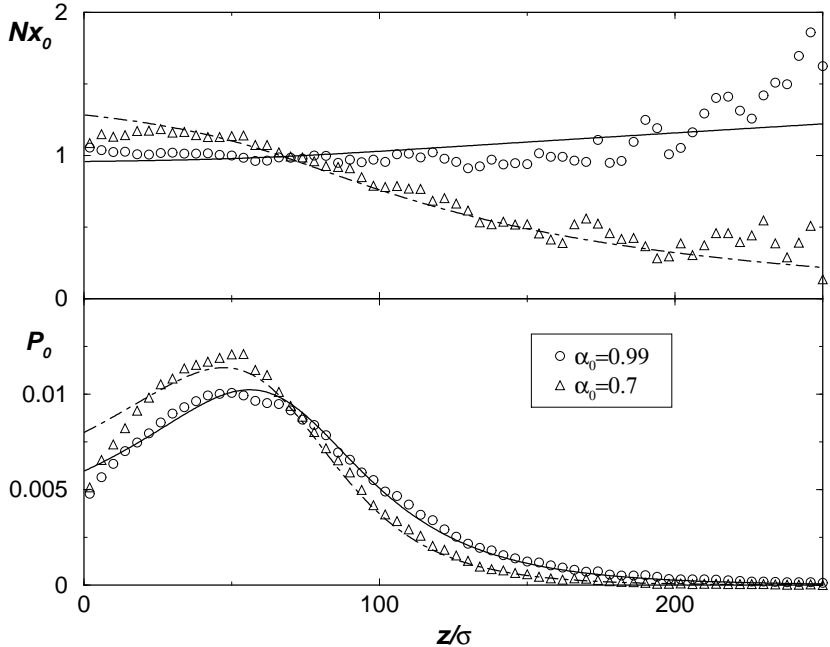}
\caption{The same as in Fig.\ \ref{fig7}, with the only difference
that here it is $m_{0}=m$ and the two values of $\alpha_{0}$
considered are $0.99$ and $0.7$, as indicated. \label{fig8}}
\end{figure}

\begin{figure}
\includegraphics[scale=0.5,angle=-90]{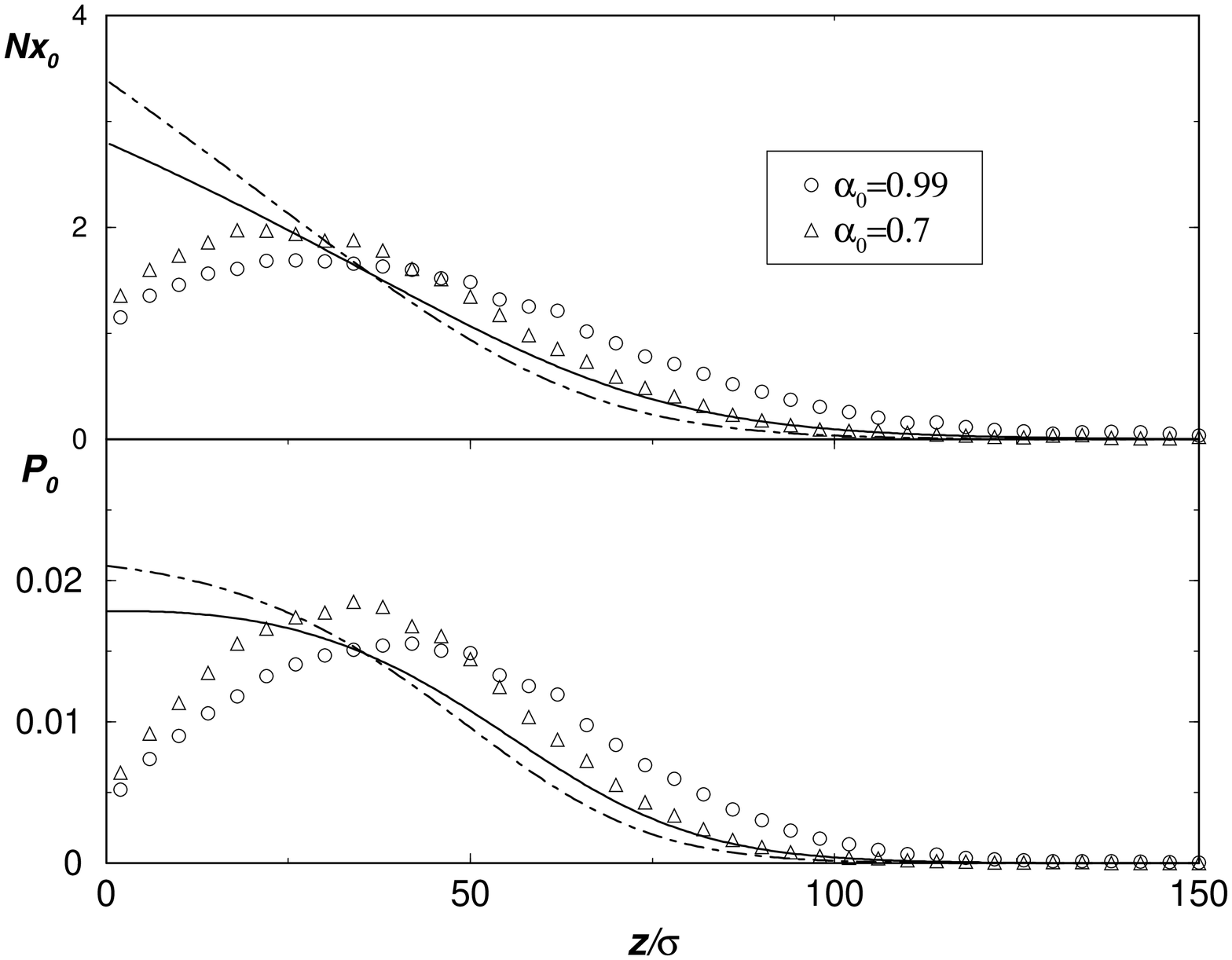}
\caption{The same as in Fig.\ \ref{fig8}, with the only difference
that here it is $m_{0}=3m$. \label{fig9}}
\end{figure}

In Figs. \ref{fig7}-\ref{fig9} the theoretical and simulation
results for $Nx_{0}$ and $P_{0}(z)$ are compared for different
values of the parameters. In all cases, those of the gas are the
same as in Fig. \ref{fig1}, i.e. $\alpha=0.95$, $N=359$, and
$W=50$, and the diameter ratio is $\sigma_{0}/\sigma=1$. On the
other hand, in Fig. \ref{fig7} it is $m_{0}/m=1/2$, in Fig.
\ref{fig8}, $m_{0}/m=1$, and in Fig. \ref{fig9}, $m_{0}/m=3$.
Finally, in each of the figures results for two different values
of the coefficient of restitution $\alpha_{0}$ are reported.

For the cases reported in Fig.\ \ref{fig7} and \ref{fig8} the
agreement between theory and simulation can be considered as quite
satisfactory, given that no fitting parameter is being used when
constructing the density profile for the impurity from the
hydrodynamic profiles of the gas. On the other hand, for
$m_{0}=3m$, Fig.\ \ref{fig9} shows strong discrepancies between
theory and simulation results. There are several possible causes
for it. One is that the heavy intruder is close, on the average,
to the vibrating wall and in this region the hydrodynamic
description for it is not accurate, as indicated in connection
with Fig.\ \ref{fig3}. In fact, the theoretical prediction for
$P_{0}(z)$ is maximum at the wall. Another closely related
possible origin of the discrepancy, is that we are assuming that
the intruder does not affect the hydrodynamic profiles of the gas.
This may be a bad approximation when the intruder is too massive
as compared with the gas particles, specially taking into account
that the number of the latter is not very large. We remind that
this number can not be increased too much, in order to keep the
system fluidized with low density and avoiding the transversal
instability mentioned in Sec.\ \ref{s3}.

We have investigated also the density profile of the intruder for
other values of the diameter ratio $\sigma_{0}/\sigma$, namely
$1/2$ and $2$, and obtained similar results.

\begin{figure}
\includegraphics[scale=0.5,angle=-90]{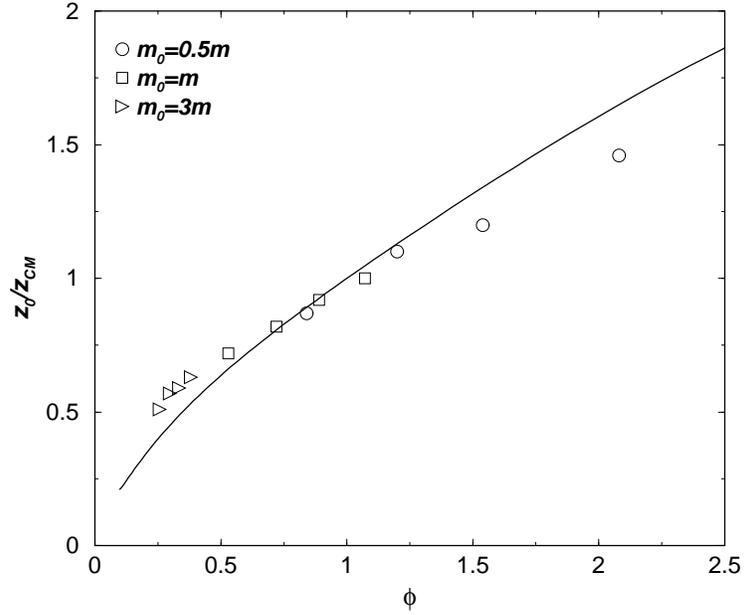}
\caption{Ratio between the impurity and the gas center of mass
positions vs the ratio of the mean square velocities $\phi$ for
systems with $\sigma_{0}= \sigma/2$. The solid line is the
theoretical prediction discussed in the text and the symbols MD
simulation results. The same symbol is used for simulations
differing only on the value of the restitution coefficient
$\alpha_{0}$, which has been varied between $0.3$ and $0.95$. The
values of the other parameters are given in the text.
\label{fig10}}
\end{figure}

\begin{figure}
\includegraphics[scale=0.5,angle=-90]{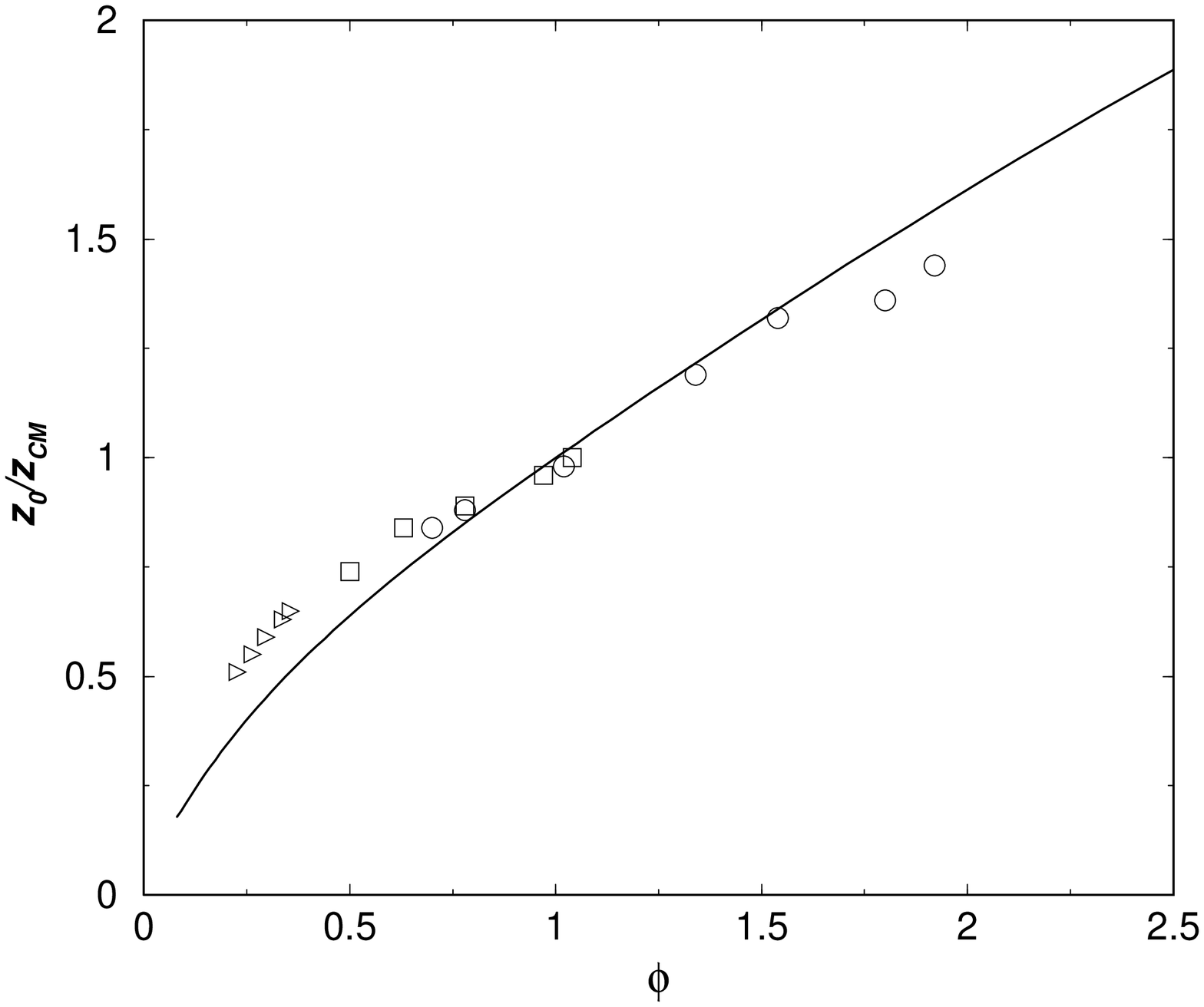}
\caption{The same as in Fig.\ \ref{fig10}, with the only
difference that here it is $\sigma_{0}=\sigma$. \label{fig11}}
\end{figure}

\begin{figure}
\includegraphics[scale=0.5,angle=-90]{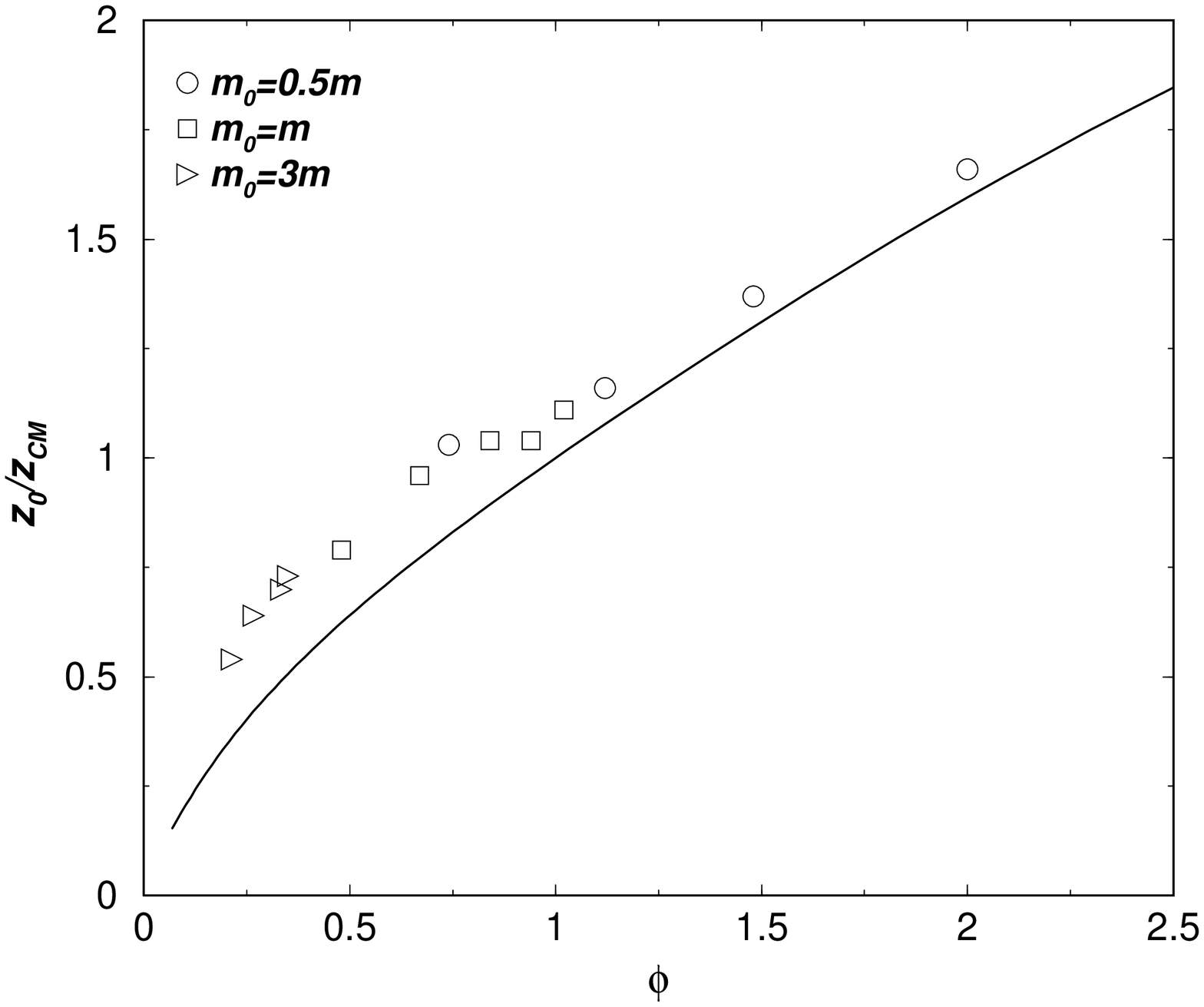}
\caption{The same as in Fig.\ \ref{fig11}, with the only
difference that here it is $\sigma_{0}=2\sigma$. \label{fig12}}
\end{figure}

>From $P_{0}(z)$, the average height of the intruder $z_{0}$ can be
computed through
\begin{equation}
\label{3.40} z_{0}= \int_{0}^{\infty} dz\, z P_{0}(z).
\end{equation}
The ratio $z_{0}/z_{cm}$, where $z_{cm}$ is the height of the
center of mass of the gas, provides a measurement of the relative
position of the intruder with respect to the gas. This ratio is
plotted in Fig.\ \ref{fig10} as a function of $\phi$ for several
systems with $\alpha=0.95$ and $\sigma_{0}/\sigma=1/2$ ($\beta
\simeq 0.0230$). The solid line is the theoretical prediction and
the symbols simulation results for different values of the mass
ratio and $\alpha_{0}$. Figures \ref{fig11} and \ref{fig12} show
the same function but for $\sigma_{0}/\sigma=1$ ($\beta \simeq
0.0172$) and $\sigma_{0}/\sigma=2$ ($\beta \simeq 0.0115$).

The agreement between theory and simulation is good in all cases,
specially in the region close to $\phi=1$. In fact, for
$m_{0}/m=3$ it is better than it could be expected from Fig.
\ref{fig9}. This good agreement is not surprising, since we are
looking now to the simplest global property of the distribution
$P_{0}(z)$. On the other hand, the simulation results for
$\sigma_{0}= 2\sigma$ are systematically above the theoretical
prediction. In general, it seems that the slope of the simulation
results is smaller than the theoretical prediction, although we
have no explanation for it. A relevant conclusion to be reached
from the above results is the dominant role played by the value of
$\phi$ in determining the relative position of the intruder with
respect to the gas. In an elastic system, $\phi$ is equal to the
mass ratio $m/m_{0}$, but in a granular gas $\phi$ can be larger
or smaller than the mass ratio due to the lack of energy
equipartition. Consequently, the relative position of an intruder
in a granular gas depends in a rather complicated way on the
coefficients of restitution of the system, the mass ratio, and the
size ratio.

\section{Conclusions}
\label{s5} In this work, we have studied the hydrodynamic profiles
of an impurity immersed in a low density granular gas that is in
an arbitrary state. Analytical predictions for the temperature and
density profiles have been derived and compared with molecular
dynamics simulations for the particular case of a vibrated system.
A satisfactory qualitative and quantitative agreement has been
found, although the discrepancies increase  as the difference
between the masses of the gas particles and the intruder
increases. When this is the case, the boundary layer next to the
vibrating wall becomes rather wide and, moreover, the intruder
influences the density profile of the gas, contrary to the tracer
limit assumed in the theory.

The theoretical study is based on the validity of the usual
Chapman-Enskog procedure to generate the hydrodynamic equations of
the system when starting from the kinetic (Boltzmann) equations.
Then, although the results are quite general and can be applied to
a variety of situations as the specific one considered here, some
caution is needed for other particular cases. For instance, in the
so-called steady simple shear state, there is an intrinsic
coupling between inelasticity and gradients \cite{SGyN96,SGyD04}.
This implies that the shear rate and the coefficient of
restitution can not be considered as independent quantities, and
the usual Chapman-Enskog procedure becomes inaccurate. This is in
contrast with the situation considered here, in which the
gradients are controlled by the vibrating wall and the external
gravitational field. In general, each physical state should be
checked to verify whether a direct expansion in the gradients is
legitimate.

For the ratio between the local temperatures of the intruder and
the gas, the theory predicts that it is constant in the bulk of
the system. This has been confirmed by the MD simulations. The
same behavior has been found in experiments \cite{FyM02,WyP02} and
MD simulations \cite{ByT02a} of vibrated systems, outside the
dilute and tracer limits. Nevertheless, an extension of the
analysis presented in this paper \cite{ByRtbp}, indicates that the
ratio depends also on the local densities ratio. What happens is
that this dependence is rather weak and gives corrections inside
the statistical uncertainties for the values of the parameters
considered in the above mentioned works.

A qualitative discussion of the relative position of the intruder
with respect to the gas was given in \cite{BRyM05}, where the
crucial role played by the non-equipartition of energy was
discussed. More precisely, it was shown that the intruder was
above or below the center of mass of the gas depending on the sign
of the pressure diffusion coefficient $D_{p}$, that in turn is
mostly determined by the sign of $\phi -1$, as seen in Eq.\
(\ref{3.32}). This is in agreement with the simulation results
reported in  Figs. \ref{fig10}-\ref{fig12}. It can be concluded
that the difference in temperature of the components of a mixture
must be taken into account when studying segregation phenomena,
i.e. the demix of the components when shaken. Of course, at higher
densities and finite volume fraction of the components, other
mechanisms \cite{DRyC93,HQyL01,JyY02} are also important and even
dominant.

It is worth to mention that the work reported here provides
another example of the accuracy of hydrodynamics to describe
granular systems under rather extreme conditions of inelasticity.
Although the inelasticity of the gas has been kept small in order
to avoid the transversal instability, predicted by the
hydrodynamic theory itself, the coefficient of inelasticity for
the gas-impurity collisions has been varied over a wide range
(typically between $0.3$ and $0.99$) and a reasonable good
agreement between theory and simulation has been found.

\begin{acknowledgments}
This research was supported by the Ministerio de Educaci\'{o}n y
Ciencia (Spain) through Grant No. FIS2005-01398 (partially
financed by FEDER funds).
\end{acknowledgments}

\appendix*

\section{Hydrodynamic profiles of the gas}
In ref.\ \cite{BRyM01}, analytical expressions for the density and
temperature profiles of an open vibrated granular gas in the
steady state were derived. They are given by
\begin{equation}
\label{a.1} n(\xi)= \frac{ m g_{0} \xi^{1+2\gamma}}{C_{d}
\sigma^{d-1} \sqrt{a(\alpha)} \left[ AI_{\gamma}(\xi)+BK_{\gamma}
(\xi) \right]^{2}}\, ,
\end{equation}
\begin{equation}
\label{a.2} T^{1/2}(\xi)=\left[ AI_{\gamma}(\xi)+BK_{\gamma} (\xi)
\right] \xi^{-\gamma},
\end{equation}
where $C_{d}=2 \sqrt{2}$ for $d=2$ and $C_{d}=\pi \sqrt{2}$ for
$d=3$, $A$ and $B$ are constants to be determined from some
boundary conditions, and $I_{\gamma}$ and $K_{\gamma}$ denote the
modified Bessel functions of first and second kind, respectively.
The dimensionless length scale $\xi$ is defined by
\begin{equation}
\label{a.3} \xi = \sqrt{a(\alpha)} \int_{z}^{\infty} \frac{d
z^{\prime}}{\lambda (z^{\prime})}\, ,
\end{equation}
with $\lambda(z) \equiv (C_{d} n \sigma^{d-1})^{-1}$ being the
local mean free path. Finally, $a(\alpha)$ and $\gamma(\alpha)$
are functions of the coefficient of normal restitution, both
vanishing in the elastic limit $\alpha \rightarrow 1$. Their
explicit expressions are given in \cite{BRyM01}. Note that the
maximum value of $\xi$ is
\begin{equation}
\label{a.4} \xi_{0}= \frac{\sqrt{a} C_{d} \sigma^{d-1} N}{W},
\end{equation}
$N$ being the total number of particles and $W$ the length or area
of the vibrating wall.

Using the hydrodynamic relation
\begin{equation}
\label{a.5} \frac{\partial p}{\partial z}=- n m g_{0},
\end{equation}
we can write $g(z)$, defined in Eq.\ (\ref{3.38}), as
\begin{equation}
\label{a.6} g(z)=-\frac{1}{T}\left( mg_{0}+ \frac{\zeta^{(0)}}{2
\nu} \frac{\partial T}{\partial z} \right) .
\end{equation}
The expression for $\partial T /\partial z$ is easily obtained by
using Eqs.\ (\ref{a.1})-(\ref{a.3}) and the properties of the
modified Bessel functions \cite{AyS65},
\begin{equation}
\label{a.7} \frac{\partial T}{\partial z}= - \frac{ 2 m g_{0}
\xi\left[A I_{\gamma +1}(\xi) -B K_{\gamma+1}(\xi)\right]}{A
I_{\gamma}(\xi)+B K_{\gamma}(\xi)}\, .
\end{equation}

In this way, we have an explicit expression for the integral
appearing in Eq.\ (\ref{3.39}) that, therefore, can be evaluated
numerically for each set of values of the parameters. Of course,
this requires to determine first the constants $A$ and $B$. In the
results to be reported here, it has been done from the simulation
data themselves, namely by fitting the position and value of the
temperature minimum \cite{BRyM01}. Another possibility that has
been proven to give accurate results would be to require the
hydrodynamic heat flux to vanish at infinite height \cite{ByR04}.
Then for a granular gas with $\alpha=0.95$, $N=359$, $v_{W}=5$,
and $W=50$, leading to $\gamma \simeq 0.021$, $a \simeq 0.010$,
and $\xi_{0} \simeq 2.065$, the values $A\simeq 5.41  $  and $B
\simeq 0.14$ are found. The solid lines in Fig.\ \ref{fig1} are
the fit to the hydrodynamic profiles of the gas obtained in this
way.

Once $A$ and $B$ are determined, it is possible to evaluate
numerically the right hand side of Eq.\ (\ref{3.39}) for each set
of values of the parameters defining the mechanical properties of
the intruder.

\end{document}